\documentclass[twocolumn,prl,preprintnumbers,superscriptaddress,showpacs]{revtex4}

\usepackage{exscale}
\usepackage{t1enc}
\usepackage{graphicx}
\usepackage{amsmath,amsfonts,amssymb,amsxtra}
\usepackage{bm}
\usepackage{textcomp}

\makeatletter

\DeclareRobustCommand{\chemical}[1]{%
  {\(\m@th
   \edef\resetfontdimens{\noexpand\)%
       \fontdimen16\textfont2=\the\fontdimen16\textfont2
       \fontdimen17\textfont2=\the\fontdimen17\textfont2\relax}%
   \fontdimen16\textfont2=2.7pt \fontdimen17\textfont2=2.7pt
   \mathrm{#1}%
   \resetfontdimens}}

\newcommand{\tbmno}{\chemical{TbMnO_3}}

\newcommand{\remno}{\chemical{RMnO_3}}

\newcommand{\TFE}{T_\text{FE}}

\newcommand{\TN}{T_\text{N}}
\newcommand{\epsb}{\varepsilon_b}

\newcommand{\kk}{\bm k}
\newcommand{\Q}{\bm Q}
\newcommand{\q}{\bm q}

\newcommand{\Pa}{\bm{P}_\text{a}}
\newcommand{\Pc}{\bm{P}_\text{c}}
\newcommand{\Ha}{\bm{H}_\text{a}}

\newcommand{\ho}{\hbar\omega}

\begin{document}

\textheight 24.85 true cm

\title{Field dependence of magnetic correlations through the polarization flop transition in multiferroic TbMnO$_3$ : evidence for a magnetic memory effect}

\author{D.~Senff}
\affiliation{II.~ Physikalisches Institut, Universit\"at zu K\"oln, Z\"ulpicher Str.~ 77, D-50937
K\"oln, Germany}

\author{P.~Link}
\thanks{Spektrometer PANDA, Institut f\"ur Festk\"orperphysik, TU Dresden}
\affiliation{Forschungsneutronenquelle Heinz Maier-Leibnitz (FRM II), TU M\"unchen, Lichtenbergstr.~ 1,
D-85747 Garching, Germany}

\author{N.~Aliouane}
\affiliation{Hahn-Meitner-Institut, Glienicker Str.~100, D-14109 Berlin, Germany}

\author{D. N.~Argyriou}
\affiliation{Hahn-Meitner-Institut, Glienicker Str.~100, D-14109 Berlin, Germany}

\author{M.~Braden}
\email{braden@ph2.uni-koeln.de} \affiliation{II.~ Physikalisches Institut, Universit\"at zu K\"oln,
Z\"ulpicher Str.~ 77, D-50937 K\"oln, Germany}

\date{\today}

\begin{abstract}
The field-induced multiferroic transition in  TbMnO$_3$ has been
studied by neutron scattering. Apart strong hysteresis, the
magnetic transition associated with the flop of electronic
polarization exhibits a memory effect: after a field sweep,
TbMnO$_3$ does not exhibit the same phase as that obtained by
zero-field cooling. The strong changes in the magnetic excitations
across the transition perfectly agree with a rotation of the
cycloidal spiral plane indicating that the inverse
Dzyaloshinski-Moriya coupling causes the giant magnetoelectric
effect at the field-induced transition. The analysis of the
zone-center magnetic excitations identifies the electromagnon of
the multiferroic high-field phase.

\end{abstract}


\maketitle

The magnetoelectric effect  couples the electric polarization $\bm P$ to an applied magnetic field $\bm
H$ and the magnetization $\bm M$ to an external electric field $\bm E$ \cite{fiebig05a,eerenstein06a}.
Very recently, giant magnetoelectric coupling has been reported for several transition-metal oxides,
such as e.\,g.~ perovskite manganites \remno~ with R=Gd, Dy, and Tb \cite{kimura03a,goto04a}, spinel
chromate \chemical{CoCr_2O_4} \cite{yamasaki06a}, spin-chain cuprate \chemical{LiCu_2O_2}
\cite{park07a}, and huebnerite \chemical{MnWO_4} \cite{heyer06a,taniguchi06a}. Common to all these
systems is a non-collinear magnetic ordering, the key element to understand the multiferroic
order\cite{katsura05a,mostovoy06a,sergienko06a}. Two non-collinear spins $\bm S_i,\bm S_j$ at a
distance $\bm r_{ij}$ induce a spontaneous electric polarization via an inverse Dzyaloshinski-Moriya
coupling
\begin{equation}\label{eq-polarization}
    \bm P\propto\bm r_{ij}\times(\bm S_i\times\bm S_j).
\end{equation}
A spiral structure with a finite angle between the spin-rotation axis $\bm S_i\times\bm S_j$ and the
propagation vector $\kk$ , therefore, induces a finite electric polarization, and a large
magnetoelectric effect may arise from  the response of the spiral to an applied magnetic field
\cite{mostovoy06a,cheong07a}.

\begin{figure}
  \includegraphics[width=0.42\textwidth]{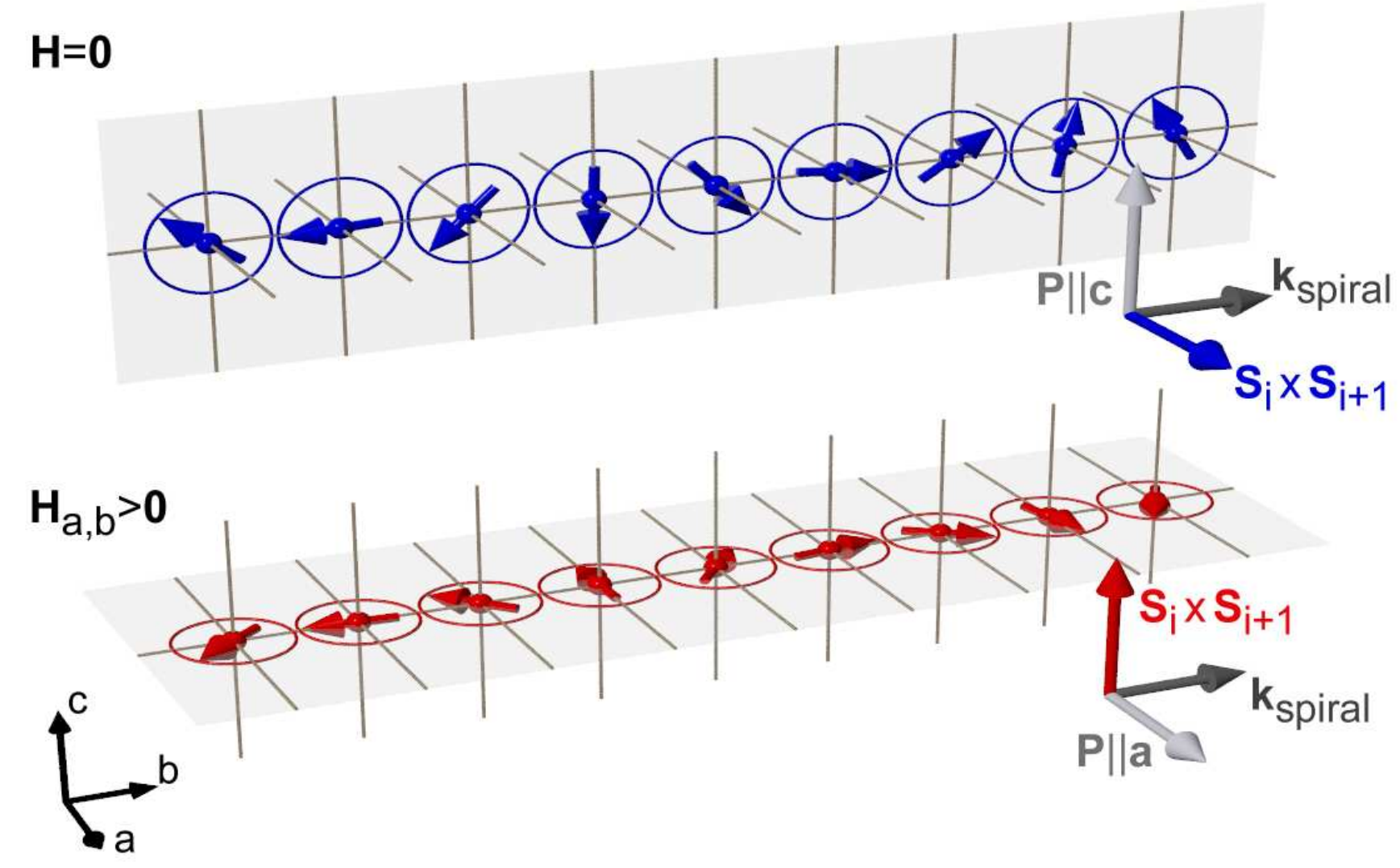}\\
  \caption{(color online) Sketch of the magnetic spiral structure in zero field inducing an electric
  polarization $\Pc$ (above). The rotation from $\Pc$ to $\Pa$ in the high-field phase is supposed
  to originate in a concomitant flop of the spiral rotation axis from $\bm a$ to $\bm c$ (below).
  }\label{Fig-Spiral}
\end{figure}

\tbmno ~ is particularly well suited for the investigation of the oxide multiferroics, as it exhibits a
rather large electric polarization and as its crystal and magnetic structure is simpler than that of
other multiferroics. \tbmno ~ crystallizes in the orthorhombic symmetry, space group $Pbnm$
\cite{alonso00a}, and the Mn magnetic moments order below $\TN$=42\,K in a collinear spin-density wave
(SDW) with incommensurate (IC) propagation vector $\kk_\text{Mn}\approx(0\,0.28\,0)$
\cite{quezel77a,blasco00a}. In a second magnetic transition at $\TFE$=28\,K, the Mn magnetic order
continuously transforms into an elliptical cycloid  with the spiral basal plane being parallel to the
$\bm{bc}$ plane (i.\,e.~ the spins rotate around the third crystallographic axis $\bm{a}$)
\cite{kenzelmann05a}. According to the general concept of eq.~ \ref{eq-polarization}, the spiral order
implies a spontaneous electric polarization $\Pc$ along $\bm c$, which is observed experimentally below
$\TFE$ \cite{kimura03a}, rendering \tbmno~ multiferroic, see Fig.~ \ref{Fig-Spiral}. Applying a
magnetic field, giant magnetoelectric effects are observed. A critical field $H_c$=7\,T along $\bm c$
completely suppresses the electric polarization \cite{kimura05a,argyriou07a}, while a field of 10.5\,T
(6\,T) parallel to $\bm a$ ($\bm b$) flops the electric polarization by $90^\circ$ from $\Pc$ to $\Pa$
\cite{kimura03a}. Since the magnetic modulation in the FE high-field phases remains along [0\,1\,0],
the $90^\circ$-rotation of the electric polarization suggests the flop of the spiral rotation axis from
$\bm{a}$ to $\bm{c}$, see Fig.~ \ref{Fig-Spiral} \cite{mostovoy06a,cheong07a}. However, a proof that
inverse Dzyaloshinski-Moriya coupling according to equation (1) accounts for the giant magnetoelectric
effects at the field-induced transition is still missing.

Measuring selected superstructure reflections by neutron or x-ray diffraction techniques
\cite{argyriou07a,arima05a,aliouane06a} it has been demonstrated, that a magnetic field $\bm H$
parallel to $\bm c$ melts the IC ordering and stabilizes a paraelectric phase with simple AFM ordering
\cite{argyriou07a}. The flop of the electric polarization for a field along $\bm a$ or $\bm b$
coincides with a first order transition into a high-field commensurate (HF-C) ordered phase with
propagation vector $\q_\text{Mn}^{HF}=(0\,0.25\,1)$ \cite{arima05a,aliouane06a}.

The mechanism in eq.~ (1)  provides for a close coupling between the dielectric properties and
magnetism which leads to hybridized phonon-magnon excitations
\cite{katsura05a,mostovoy06a,sergienko06a} as they were proposed long ago \cite{smolenskii83a}. While
such $\it electromagnon$ was proposed on the basis of Infrared (IR) optical-spectroscopy on TbMnO$_3$
\cite{pimenov06a}, only the correspondence of the peaks in the IR spectrum with neutron scattering
measurements of the IC zone-center magnetic excitations \cite{senff07a} documents the mixed
phonon-magnon character of the excitation. This rather soft electromagnon mode possesses the correct
symmetry to rotate the spiral plane from the $\bm{bc}$ to the $\bm{ab}$ plane and appears to be
activated when field is applied along $\bm a$ or $\bm b$.

In this work we follow the antiferromagnetic zone-center
excitations through the polarization flop transition with $\bm H$
applied parallel to $\bm a$ using inelastic neutron scattering.
Our measurements show pronounced changes in the magnetic
excitations at the critical field, while analyses of the spectra
are consistent with a magnetic spiral where the magnetization
rotates within the $\bm{ab}$ plane as predicted by the inverse
Dzyaloshinski-Moriya coupling. Detailed measurements of the
propagation vector $\kk$ as a function of magnetic field reveal a
magnetic memory effect that is driven by domain wall locking.

Neutron scattering experiments were performed at the cold triple-axis spectrometer PANDA (FRM-II,
Garching) using the same single crystal as in our previous studies \cite{senff07a}. The sample was
mounted with the scattering plane defined by [0\,1\,0] and [0\,0\,1] in a 15\,T-cryomagnet and the
field applied along the vertical $\bm a$-axis. The $Pbnm$-lattice constants are $a=5.302\,\text{\AA}$,
$b=5.857\,{\text{\AA}}$, and $c=7.402\,{\text{\AA}}$, which in the following will be used to index all
vectors of reciprocal space. Monochromatic neutrons were selected and analyzed using the
(0\,0\,2)-Bragg reflection of pyrolytic Graphite (PG). In the standard set-up the energy of the
analyzed neutrons was fixed to $E_f$=4.66\,meV $(k_f=1.50\,{\text{\AA}^{-1}})$, but in order to enhance
the experimental resolution selected scans were repeated with $E_f$=2.98\,meV
$(k_f=1.20\,{\text{\AA}^{-1}})$.

\begin{figure}[t]
  \includegraphics[width=0.425\textwidth]{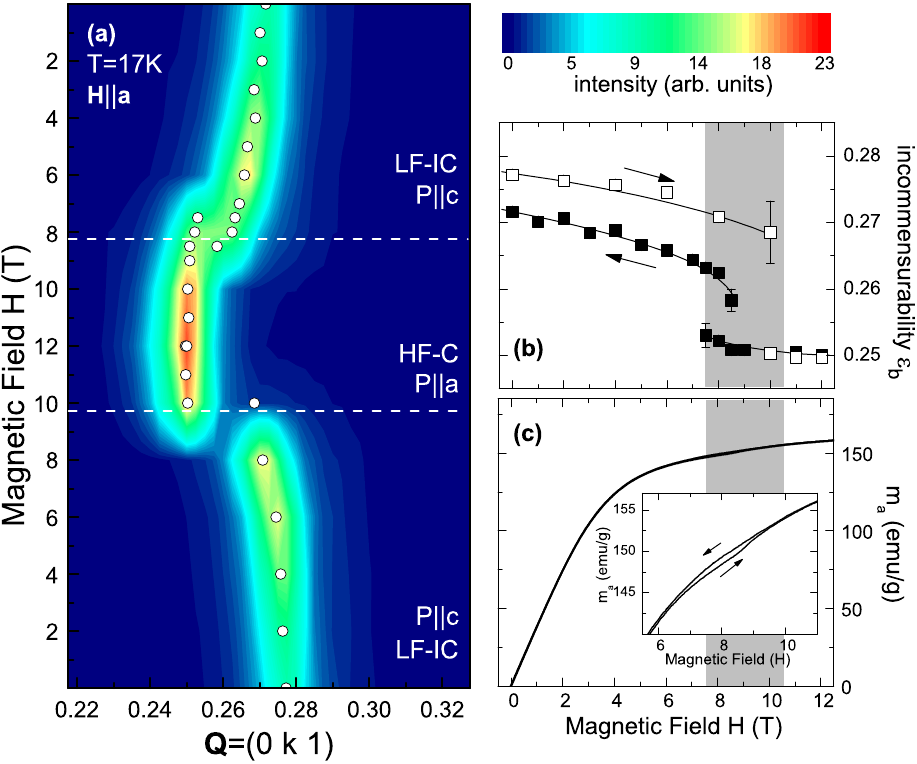}\\
  \caption{(color online) (a) Cut through reciprocal space along $\Q=(0\,k\,1)$ for a complete field cycle
  $H_a=0\,\text{T}\rightarrow12\,\text{T}\rightarrow0\,\text{T}$ and $\Ha||\bm a$. Open points mark
  the positions of the magnetic superstructure reflections, dashed lines denote the phase transitions between the
  LF-IC and the HF-C phase as determined in Ref.~ \onlinecite{meier07a}. (b)  Field dependence of the incommensurability $\epsb$ for the complete field-cycle. The grey shaded
  regions mark regimes with coexistence of both magnetic phases, solid lines are included as guides to
  the eye. (c) Hysteresis of the magnetization measured with  $\Ha||\bm
  a$ sensing essentially the alignment of Tb-moments, note that
  there is a minor hysteresis at the LF-IC to HF-C transition.
  }\label{Fig-Map-ICtoC}
\end{figure}

Fig.~ \ref{Fig-Map-ICtoC}  shows the response of the magnetic superstructure reflection
$\Q=(0\,\epsb\,1)$ to a magnetic field $\Ha||\bm a$ for a complete field cycle
$H_a=0\,\text{T}\rightarrow12\,\text{T}\rightarrow0\,\text{T}$ in the ferroelectric regime at
$T=17\,\text{K}$. The intensity of the magnetic reflection increases continuously with increasing
field, see Fig.~ \ref{Fig-Map-ICtoC}a. In contrast, up to 8\,T the magnetic modulation described by the
zero-field incommensurability $\epsb^{0}=0.2772(1)$ remains nearly unchanged. With further increase of
the field across the critical value $H_{\uparrow}^a\simeq9\,\text{T}$, $\epsb$ exhibits a sharp
discontinuity, and for $H>H_{\uparrow}^a$ the magnetic intensity is centered at the commensurate
position $\Q=(0\,0.25\,1)$, see Fig.~ \ref{Fig-Map-ICtoC}a and b. In the vicinity of $H_{\uparrow}^a$,
a region of coexistence of low-field incommensurate (LF-IC) and HF-C phases exists, and the transition
is clearly of first order, in good agreement with recent scattering experiments
\cite{arima05a,aliouane06a} and thermodynamic studies \cite{meier07a}. On the downward run a pronounced
hysteresis is observed, as the system switches back into the LF-IC phase at
$H_{\downarrow}^a\simeq8\,\text{T}$ \cite{meier07a}. Again, around $H_{\downarrow}^a$ a region of
coexistence of both phases is found, which is broader than that in the upward stroke. Furthermore, upon
reentering into the LF-IC phase the magnetic incommensurability $\epsb$ does not recover its initial
value, see Fig.~ \ref{Fig-Map-ICtoC}b. Switching back from the HF-C phase we find
$\epsb(8\,\text{T}_{\downarrow})=0.2624(3)$, which is significantly smaller than in the upward run,
$\epsb(8\,\text{T}_{\uparrow})=0.2708(1)$. At zero field $\epsb^{0\downarrow}=0.2716(1)$ is smaller
than the initial value $\epsb^0$ determined before the field sweep. We can only recover the original
zero-field incommensurability by heating the system up into the paraelectric SDW-phase and subsequent
re-cooling it to 17\,K. In the FE spiral phase, \tbmno~ memorizes its magnetic history as the size of
the incommensurability $\epsb^0$ at zero field depends on the magnetic diary of the sample.
Measurements of magnetization from a single crystal of \tbmno~ with $H\| a$ show that there is no large
effect that tracks the hysteresis of the magnetic wave vector as we found in our neutron measurements.
The magnetization hysteresis that is associate with the transition to the HF-C phase is very small, see
Fig.~ \ref{Fig-Map-ICtoC}c. This indicates that the ferromagnetic alignment of Tb-spins that occurs
with applied magnetic field is of little consequence to the memory effect we find here. A more
plausible explanation is that the transition into the HF-C phase induces domain walls which at least
partially remain when switching back to the LF-IC phase. These domain walls are only suppressed when
fully leaving the spiral phase upon heating \cite{note1}. Memory effects at combined spin and
polarization flop transitions should be of general importance in view of future applications in
magnetoelectric data storage.

\begin{figure}[t]
  \includegraphics[width=0.42\textwidth]{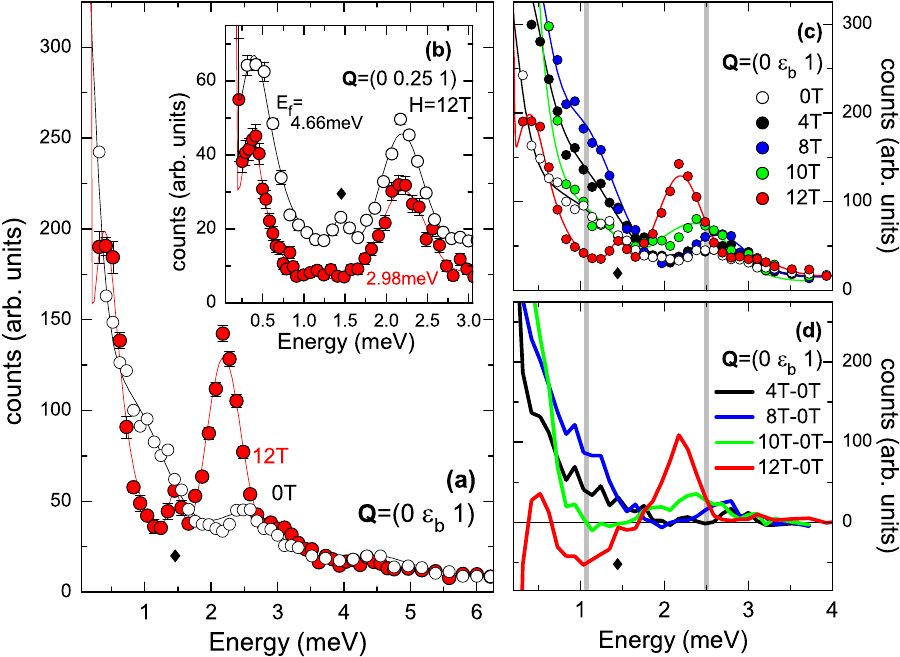}\\
  \caption{(color online) (a) Energy scans at the magnetic zone-center $\Q=(0\,\epsb\,1)$ in the
  LF-IC phase at $H=0\,\text{T}$ and in the HF-C phase at $H=12\,\text{T}$, both recorded with
  $E_f=4.66\,\text{meV}$. The zero-field data are the same as shown in Ref.~ \onlinecite{senff07a}. (b)
  High-resolution data of the 12\,\text{T}-spectrum for $E_f=2.98\,\text{meV}$. (c) Raw-data scans for
  various magnetic fields, and (d) difference spectra derived from the data presented in (c). Solid lines
  correspond to fits to the data in (a)-(c), vertical bars mark the position of the magnetic excitations
  at zero field, small diamonds label regions with spurious contributions in the 12\,\text{T}-spectra.
  }\label{Fig-Magnon-Field}
\end{figure}

We now turn to the field dependence of the magnetic excitation spectrum. Inelastic scans at the
magnetic zone-center $\Q=(0\,\epsb\,1)$ for $T=17\,\text{K}$ are presented in Fig.~
\ref{Fig-Magnon-Field}. At zero field, several magnetic signals are detected. The feature around
4.5\,meV corresponds to a crystal field (CF) excitation in the Tb-subsystem \cite{senff07a}. In the
following discussion we will ignore this CF-excitation, as up to the maximum field investigated,
$H=12\,\text{T}$, no significant changes can be observed in the spectra above $\approx4\,\text{meV}$,
see Fig.~ \ref{Fig-Magnon-Field}a. In the regime of the Mn spin-wave excitations below 4\,meV the
impact of the magnetic field is, in contrast, remarkable, see Fig.~ \ref{Fig-Magnon-Field}c and d.

At zero field, the magnon spectrum consists of three different low-energy branches: One mode with very
low energy, and two modes at finite energies, 1.0\,meV and 2.5\,meV, respectively, are observed at the
magnetic zone center \cite{senff07a}. With increasing field, this characteristic structure of the
spectrum remains unchanged up to $H_a=8\,\text{T}$, and the data for 4\,T and 8\,T can well be
described using similar parameter sets as those for the zero-field data. The observed spectral weight
of all modes increases following the elastic signal with increasing field, but for $H_a\leq8\,\text{T}$
we do not observe significant changes in the spin-wave frequencies. Note, that the broad response close
to $\omega=0\,\text{meV}$, which has to be ascribed to the low-lying spin-wave mode, can not be fully
separated from the elastic signal.

With the transition into the HF-C phase the spectrum exhibits prominent changes, see Fig.~
\ref{Fig-Magnon-Field}d. The signal around 1.0\,meV is completely suppressed, and for
$H^{a}=12\,\text{T}$ the spectrum can be decomposed into two intense excitations centered around
0.5\,meV and 2.25\,meV, and a weaker feature around 3.0\,meV. A fourth, rather sharp signal around
$\approx1.5\,\text{meV}$, marked by small diamonds in Fig.~ \ref{Fig-Magnon-Field} can be ascribed to a
spurious contribution \cite{note2}. The low-energy spin-wave spectrum of the HF-C phase consists  of
three branches, similar to the spectrum in the LF-IC phase \cite{senff07a}.

The data measured at 10\,T-data can  be described by a weighted summation of the $8\,\text{T}$ and
$12\,\text{T}$ spectra, see Fig.~ \ref{Fig-Magnon-Field}c, consistent with the co-existence regime of
the C and IC wave vectors shown in Fig. 2b.

\begin{figure}[b]
  \includegraphics[width=0.29\textwidth]{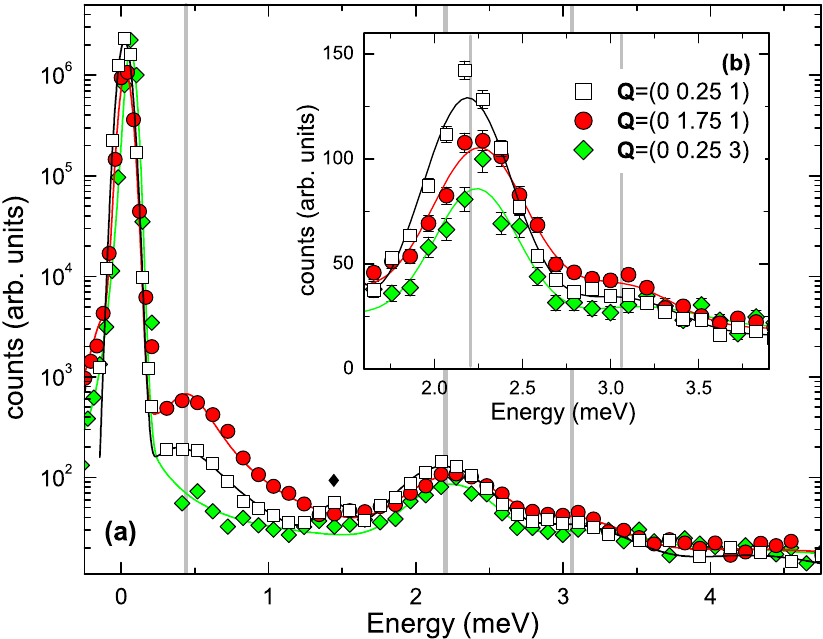}\\
  \caption{(color online) (a) Comparison of raw-data energy scans in the HF-C phase for three equivalent
  $\Q$-positions on a logarithmic scale, and (b) on a linear scale. Solid lines denote fits to the data
  and grey vertical bars mark the HF magnon-frequencies. Small black diamonds denote regions with spurious
  contributions for $\Q=(0\,0.25\,1)$.
  }\label{Fig-Q-Geometry}
\end{figure}

In the zero-field phase with $\bm P||\bm c$ longitudinal polarization analysis allowed for the
unambiguous identification of the phason mode and of the Goldstone Boson of the FE spiral structure
\cite{senff07a}. The applied magnetic field prohibits a similar polarization analysis for the $\Pa$
HF-C phase. However, the systematic survey of the $\Q$-dependence of the magnon signal can determine
the character of the different magnon branches in the HF-C phase, as only magnetic polarization
perpendicular to the scattering vector, $\bm S_\bot$, contributes to the neutron scattering intensity.
In Fig.~ \ref{Fig-Q-Geometry} we compare the excitation spectra recorded at the three different
$A$-type zone centers $\Q_1=(0\,0.25\,1)$, $\Q_2=(0\,1.75\,1)$ and $\Q_3=(0\,0.25\,3)$. At these three
equivalent $\q$-points, the observed magnon frequencies are identical, $\ho_1=0.44(1)\,\text{meV}$,
$\ho_2=2.20(2)\,\text{meV}$ and $\ho_3=3.06(2)\,\text{meV}$. However, significant changes can be
observed in the distribution of spectral weight among the three modes. The mode $\omega_2$ is strongest
for $\Q=\Q_1$, whereas the branches $\omega_1$ and $\omega_3$ are most intense at $\Q_2$ and are
remarkably suppressed at $\Q_3$. With $f(Q)$ denoting the magnetic form factor and $\alpha$ denoting
the angle between $\Q$ and the magnetic polarization, the observed intensity is given by
\begin{equation}\label{Eq-intensity}
 I \propto f^2(Q)\sin^2(\alpha).
\end{equation}

With the three chosen $\Q$ points the geometrical conditions
significantly vary as the angle with $\bm c$, $\alpha _c$,
changes : $\alpha_{c1}=17.5^\circ$ at $\Q_1$,
$\alpha_{c2}=65.5^\circ$ at $\Q_2$, and $\Q_3$ is almost parallel
to $\bm c$, $\alpha_{c3}=6^\circ$. The strong suppression of
spectral weight for $\omega_1$ upon rotating the scattering
vector towards $\bm c$ thus implies a magnetic fluctuation
polarized predominately along $\bm c$. Indeed, from the observed
intensities we find
$I^\text{obs}_{\omega_1,\Q_1}:I^\text{obs}_{\omega_1,\Q_2}=1:4.8(4)$,
which is close to the expected value for an entirely $\bm
c$-polarized mode, $I^\text{c}_{\Q_1}:I^\text{c}_{\Q_2}=6.2$. The
second mode $\omega_2$ exhibits opposite behavior.  The loss of
scattering intensity from $\Q_1$ to $\Q_3$ is fully explained by
the magnetic form factor $f(Q)$, implying that $\omega_2$ is
polarized perpendicular to $\bm c$. The third mode $\omega_3$,
although by far the weakest of the three magnon excitations,
exhibits similar behavior as $\omega_1$ with the most intense
signal for $\Q=\Q_2$, see Fig.~ \ref{Fig-Q-Geometry}b, and we may
thus conclude that $\omega_3$ also possesses a dominant component
along $\bm c$.

The magnetic structure in zero field is described by a magnetic cycloid in the $\bm{bc}$-plane
\cite{kenzelmann05a}, and it has been shown both theoretically and experimentally that the low-energy
spin-wave spectrum consists of three branches with different character \cite{senff07a,katsura07a}. One
of the modes is the sliding mode of the spiral and is polarized within the spiral plane. The two other
modes correspond to magnetic fluctuations perpendicular to the spiral plane and are polarized along
$\bm a$. These modes rotate the spiral basal plane around $\bm c$ and around $\bm b$, respectively. The
latter one is identified as the electromagnon observed in IR spectroscopy \cite{pimenov06a}. This mode
thus possesses the symmetry to rotate both the magnetic spiral plane and the ferroelectric
polarization.

Applying the arguments of the zero-field analysis to the spectrum of the HF-C phase, our observations
are fully consistent with the field-induced flop of the spiral to the $\bm{ab}$-plane, as the
polarization patterns of the spin-wave excitations are rotated by $90^\circ$. One of the three modes in
the HF-C phase, $\omega_2$, is polarized within $\bm{ab}$, and this mode should correspond to the
sliding mode of the commensurate spiral. Compared to the LF-IC phase the phason energy is strongly
enhanced, increasing from $\approx\!0.2\,\text{meV}$ for $H=0\,\text{T}$ to 2.20\,meV in the HF-phase.
The large phason energy indicates a strong pinning in the HF-C phase resulting in a very anharmonic
magnetic modulation associated with domain boundaries. This further corroborates our interpretation
that the magnetic memory effect arises from domain boundaries introduced in the HF-C phase.
Furthermore, the magnetic field will additionally trap the phase of the spiral. The two other HF-modes
are both polarized predominantly along $\bm c$, i.\,e.~ perpendicular to the flopped spiral plane.
These branches, thus, correspond to the two $\bm a$-modes of the zero-field spiral, and following the
above argumentation, at least one of these modes is expected to couple strongly to an alternating
electric field along $\bm c$ and should be visible in optical spectroscopy. This mode is the
multiferroic electromagnon, but our neutron scattering experiment cannot determine which of the two
$\bm c$-modes at high field corresponds to the electromagnon. We are,unfortunately, not aware of an
optical study to observe the electromagnetic response of TbMnO$_3$ in the HF-C phase.

In conclusion, we have studied the field dependence of elastic and inelastic neutron scattering in
multiferroic \tbmno . The magnetic superstructure reflections exhibit strong hysteresis across the
magnetic transition accompanying the flop of the electronic polarization and, most interestingly, there
is a clear magnetic memory effect. After the field sweep at low-temperature, the system is not found in
the same magnetic phase as the one obtained by zero-field cooling. Such memory effect should be a
generic feature at incommensurate to commensurate spin-flop transitions in spiral magnets. The magnetic
excitation spectrum at the zone center in the commensurate high-field phase consists of three different
modes, whose polarization patterns were determined by examining different $\Q$-positions. These results
are fully consistent with the assumed field-induced flop of the spiral plane from $\bm{bc}$ to
$\bm{ab}$ plane, which explains the giant magnetoelectric effect arising from the rotation of the
electric polarization from $\Pc$ to $\Pa$. One of the two observed $\bm c$-polarized modes is the
electromagnon of the multiferroic high-field phase. These results indicate that the inverse
Dzyaloshinski-Moriya coupling, see equation (1), which fully explains the temperature driven
multiferroic coupling, also accounts for the field-induced giant magneto-electric effect.

{\it Acknowledgments} This work was supported by the Deutsche
Forschungsgemeinschaft through the Sonderforschungsbereich 608.
We thank D.~ Etzdorf for his technical support with the magnet.



\begin{thebibliography}{25}
\bibitem{note1} {All inelastic data
have been recorded in an upward run of the magnetic field after
checking $\epsb^0$ at zero field.}
\bibitem{note2} {This signal disappears upon changing the wavelength, see
Fig.~ \ref{Fig-Magnon-Field}b, it is not observed at equivalent
$\Q$-positions, and it exhibits no significant dispersion.}
\end{thebibliography}

\end{document}